# The Meat of the Matter: A thumb rule for scavenging dogs?

Anandarup Bhadra[1], Debottam Bhattacharjee[1], Manabi Paul[1] and Anindita Bhadra[1*]

[1]Behaviour and Ecology Lab, Department of Biological Sciences, Indian Institute of Science Education and Research-Kolkata, West Bengal, India

***Corresponding author:**

Behaviour and Ecology Lab,

Department of Biological Sciences**,**

Indian Institute of Science Education and Research – Kolkata

P.O. BCKV Main Campus, Mohanpur,

Nadia PIN 741252,

West Bengal, INDIA

*tel.* 91-33-25873119

*fax* **+91-33-25873020**

*e-mail:* abhadra@iiserkol.ac.in

**Abstract**

Animals that scavenge in and around human localities need to utilize a broad range of resources. Preference for any one kind of food, under such circumstances, might be inefficient. Indian free-ranging dogs, *Canis lupus familiaris* are scavengers that are heavily dependent on humans for sustaining their omnivorous diet. The current study suggests that because of evolutionary load, these dogs, which are descendents of the decidedly carnivorous gray wolf, still retain a preference for meat though they live on carbohydrate-rich resources. The plasticity in their diet probably fosters efficient scavenging in a competitive environment, while a thumb rule for preferentially acquiring specific nutrients enables them to sequester proteins from the carbohydrate-rich environment.

## Introduction

Dogs (*Canis lupus familiaris*) are thought to have evolved from gray wolves (*Canis lupus lupus*) in East Asia about 15000 years ago with multiple founding events(Clutton-Brock 1995; Cohn 1997; Vila 1997; Savolainen et al. 2002). There is much controversy over the exact time of origin as well as the precise path of evolution (Pennisi 2002) of modern dogs which are currently recognized, not as a distinct species, but as a subspecies of wolves. Yet, behaviourally dogs are significantly different from their ancestors, the gray wolves (Miklosi 2007). Much of these differences may have arisen because of domestication, while some could have actually driven the process of evolution from the wolves to the dogs (Trut 1999; Axelsson et al. 2013). Unlike wolves, which hunt for meat and occasionally scavenge (Mech & Boitani 2003; Forbes & Theberge 1992), their modern-day descendents - the domesticated dogs are fed by their owners in controlled amounts, often leading to over feeding. Free-ranging dogs exist in many countries like Mexico (Daniels & Bekoff 1989; Ortega-Pacheco et al. 2007), Italy (Boitani 1983), Zambia (Balogh 1993), Zimbabwe (Butler et al. 2004), Sri Lanka (Matter & Wandeler 2000), India (Pal 2001; Vanak & Gompper 2009), Ecuador (Kruuk & Snell 1981), Philippines (Childs et al. 1998), Nepal and Japan (Kato & Yamamoto 2003) etc, and live almost entirely by scavenging (Vanak & Gompper 2009; Vanak et al. 2009), with occasional hunting and begging for food. This makes them an ideal model system to study the effects of the earliest form of domestication. Indian free-ranging dogs have perhaps existed as an integral part of human settlements for millennia. Their earliest mention dates back to the Mahabharatha, the Indian epic which has been dated to a period ranging from $9^{th}$ century BC to $4^{th}$ century CE (van Buitenen 1973; Debroy 2008). The dog has appeared in many ancient Indian texts and folklores over the ages, sometimes as a domesticated animal and sometimes as a stray (Debroy 2008). Dogs breed annually, and hence they have lived in their current state in India for at least 1000 generations. This should have provided enough opportunities for adapting to the scavenging lifestyle that they lead as an integral part of the human ecology today (Pal 2001).

In the absence of hunting and because of their dependence on human charity (begging for food) and leftovers (feeding in garbage dumps) free-ranging dogs don't often encounter meat. They live mostly on a carbohydrate rich omnivorous diet (biscuits, breads, rice, lentil, fish bones, occasional pieces of decomposing meat from a carcass and even mangoes, cow dung and plastic). The free-ranging dogs in India are often persecuted by humans and they live in a highly competitive environment, such that territorial fights at feeding sites are common (Das and Bhadra, in preparation). The competition over food is even translated to parent-offspring conflict over food given by humans as early as 10 weeks of the pups' age (Paul et al, under review). We wonder if they still have a strong preference for meat like the pets (Houpt et al. 1978) or whether they have lost the preference to maximize the utilization of available resources as suggested by  Thorne (1995).

## Methods

We carried out several choice experiments in which a random dog was provided with three food options simultaneously and the order of inspection and eating of the food was recorded. We call this module the One Time Multi-option Choice Test (OTMCT). Since it is expected that these dogs, living in a highly competitive environment, would eat the preferred food first, we recorded the order in which the food was consumed. The data for only those cases where all the options were at least inspected were used for analysis. The experiments were conducted in Kolkata (22°34'10.92" N, 88°22'10.92" E), Kalyani (22°58'48" N, 88°28'48" E) and Raiganj (25°37'12" N, 88°7'12" E), West Bengal, India.

In the OTMCT experiments we used small quantities of food (less than 10 ml) for each option as we did not want the quantity of food to be a stimulus for the undernourished free-ranging dogs. We used sample sizes of minimum 30, giving each dog the choice test only once. This also eliminated the effect of learning and we could get a clear representation of the preference already formed at the population level.

Absolute choice was defined as the total number of times each option was chosen in a particular experiment. Choice was taken as the complete consumption of a particular option, except in experiment 1, where both licking and consumption of an option was taken as choice. In the cases where no clear absolute choice was seen, the eating order was computed for each experiment. A 3x3 matrix was constructed with the three options in the columns and the number of times each option was chosen first, second and third respectively in the rows. Now, a contingency chi-squared test was done to determine whether the tables were random. If they were significantly different from random, then the option that was chosen first the highest number of times was taken to be chosen first at the population level, that is, the first preference. Similarly the options chosen second and third were also determined.

**(i) Experiment 1**

In this experiment we wanted to test the preference, if any, between sources of protein and carbohydrates that are known to adult free-ranging dogs, when visual cues were not present. So the options we provided in the OTMCT were B1 (bread + chicken extract); B2 (bread + water) and B3 (bread), where the 3 pieces of bread looked identical (please see OSM for details of chicken extract preparation and protein content estimation). Chicken extract contains a small percentage of proteins (less than 0.25% w/v as determined by Bradford Method (Bradford 1976)) and we wanted to see if that is enough for the dogs to prefer it over the carbohydrate rich bread. B2 is present as a control for moistness. We wanted to discount the effect of the moistness of the chicken extract in this single experiment as pet dogs are known to prefer moist food over dry food (Kitchell 1978). We used 30 adults for this experiment.

B1 was chosen significantly more often than B2 or B3 (two tailed Fisher's Exact Test; $p = 0.00$), which were chosen equally as often (two tailed Fisher's Exact Test; $p = 1$) (Figure 1) (Table 1). It seems that the dogs have a clear preference for proteins over carbohydrates. But is this preference translated to real items of food in the highly competitive natural environment when visual cues are provided?

**(ii) Experiment 2**

In this experiment the options we provided in OTMCT were A (bread); B (bread + chicken gravy) and C (cooked chicken), such that the three options were visually different. There was no significant difference in terms of absolute choice (A vs B: two tailed Fisher's Exact Test; $p = 1$, B vs C and A vs C: two tailed Fisher's Exact Test; $p = 0.49$) (Table 1). In terms of eating order these dogs clearly preferred chicken over bread soaked in chicken gravy over dry bread (chi square = 45.37, df = 4, $p = 0.00$) (Table 2) (Figure 2a). So they do prefer protein over carbohydrate when they are provided with a visually identifiable protein source (cooked chicken). But do they follow a gradient of chicken even in the absence of visual cues?

**(iii) Experiment 3**

In the third experiment all visual cues were removed by making the three options in the choice test look alike. Any preference shown here can only be because of olfactory cues. A controlled experiment was used to test the preference for increasing concentration of meat smell in the free-ranging dogs. We assumed that the addition of low quantities of chicken extract to a protein rich food source would not cause significant increase in its protein concentration. Hence we soaked Pedigree (PEDIGREE® Puppy Chicken & Milk) tablets containing 24% protein in different concentrations of chicken extract for this experiment to create a gradient of chicken smell on a base of synthetic protein-rich food. . The three options in the experiment 3A were T (Pedigree soaked in 100% chicken extract); I (Pedigree soaked in 50% chicken extract) and S (Pedigree soaked in water) and those in experiment 3B (control) were U (bread + 100% chicken extract); M (bread + 50% chicken extract) and L (bread + water). Thus, if the olfactory cues convey any information about the absolute nutritional content (let us call it the protein smell) the options in experiment 3A are not expected to differ in this context. The dogs should not, therefore, display any preference for a particular option in this experiment, if they choose food by smelling its protein content only. However, in the control, the absolute concentration of protein is expected to be different in the three choices. Thus, based on the protein smell, the dogs should display a preference for increasing concentration of chicken extract. It is possible, however, that the dogs simply prefer the smell of meat, irrespective of its nutritional value. Under such circumstances they should clearly display a preference for increasing concentrations of chicken extract in both the experiments. 30 adults each were used for the experiment and control conditions.

In terms of absolute choice there was no preference in both experiments 3A (T vs I: two tailed Fisher's Exact Test; $p = 0.71$, I vs S: two tailed Fisher's Exact Test; $p = 0.53$, S vs T: two tailed Fisher's Exact Test; $p = 0.18$) and 3B (two tailed Fisher's Exact Test; $p = 1$) (Table 1). But in both the experiments the eating order revealed a preference for higher concentrations of chicken extract (3A: chi square = 24.00, df = 4, $p = 0.00$; 3B: chi square =

11.14, df = 4, p = 0.02) (Figure 2b and 2c) (Table 2). It is not unlikely that the dogs might be using a simple thumb rule of choosing food that smells like meat over any other available option. Such a thumb rule would then dictate them to ignore sources of more nutritious food that does not smell like meat for a less nutritious food having meat smell.

**(iv) Experiment 4**

In this experiment we tested if such a thumb rule might exist by presenting an OTMCT to the dogs where meat smell and protein were given in reverse gradients to the same set of dogs. The options provided were F1 (90 pellets of pedigree soaked in water); F2 (45 pellets of pedigree and half bread in 50% Chicken extract) and F3 (One and a half bread in 100% chicken extract). The three options were mashed, and a pinch of turmeric powder was added to each of them to make them visually identical. This experiment was video recorded and analysis was carried out in the blind. The absolute protein gradient is equivalent to the pedigree gradient. 30 dogs were used for this experiment.

In terms of eating order the dogs clearly preferred F3 over F2 over F1 (chi square = 16.99, df = 4, p = 0.00), thereby confirming the presence of the thumb rule (Figure 2d) (Table 2).

## Discussion

In the first experiment, bread with chicken extract was chosen over both wet and dry bread (in terms of absolute choice) showing a preference for protein over carbohydrates. Since the moist bread was not preferred, the preference of B1 can only be because of the presence of chicken extract and not because of moistness. Unlike the pet dogs, the Indian free-ranging dogs do not seem to have any preference for moist food over dry food (Kitchell 1978). Since these dogs scavenge from dumps and human disposals, their food is likely to get dried and desiccated due to prolonged exposure to the tropical sun. Under such circumstances, an indifference to moistness of food is likely to help maximise the utilization of resources, and increase the efficiency of scavenging in the face of intense competition.

The experiment with real food provided visual cues in addition to olfactory ones, and revealed a clear preference for chicken over bread soaked in chicken gravy over bread. In fact, in some cases the choice was made even without close inspection. So, the free-ranging dogs clearly are partial to chicken in various forms (cooked chicken, chicken extract or gravy) as is apparent from the order of eating. This preference for meat (chicken being used as the predominant meat of West Bengal and most other parts of India) is consistent with that in pet dogs. Even as scavengers, the Indian free-ranging dogs seem to retain a preference for meat, in direct contradiction to Thorne (1995). This clearly does not limit the utilization of resources (since none of the options were completely rejected in the choice tests); instead it ensures the differential utilization of preferred resources. The scavenging habit is maintained by the flexibility of the diet and physiology of these dogs. But their physiology is also constrained by an evolutionary load which is the requirement for proteins (since their ancestors were predators subsisting solely on the meat of prey). Preferentially eating the meat first might be a behavioural adaptation to maximize the utilization of resources that might have any quantity of protein. Given the possibility of losing available resources to competitors, the dogs should rightly eat valuable nutritious food as soon as they locate it. But there is a clear preference for the actual meat over the gravy. This could be explained by the visual cue of the chicken, the difference in the intensity of the meat smell between the two or the nutritional difference (cooked meat has more protein than meat extract).

In experiment 3, although there was no difference in the absolute choice of the three options in both the experiments, there was a clear preference for higher concentration of chicken extract in terms of the order of eating. Moreover, there was no difference in the order of selection in the two experiments (two tailed paired t-test, df = 8, p = 0.69) showing that the dogs are indifferent to the presence of carbohydrate or protein, when they are given a gradient of meat smell. It seems that the Indian free-ranging dogs not only have a preference for meat, they actually are able to detect out higher intensities of meat smell. This is surely a great advantage for protein-starved scavengers who constantly compete for food because a higher meat smell, in nature, would mean higher quantity of meat and thus more protein and nutrition.

It appears that Indian free-ranging dogs follow a thumb rule- “always choose a food with higher intensity of meat smell first”. But this has not been clearly demonstrated in the third experiment as chicken extract does have a small amount of protein, forming a protein gradient (however slight) over and above the Pedigree in the three options in the experiment 3A. So, in a further experiment, the meat smell gradient and absolute protein gradient were reversed: F1<F2<F3 in terms of the concentrations of both chicken extract and bread, while F1> F2> F3 in terms of the concentrations of the synthetic protein as well as the absolute protein. The preference shown by the dogs is clearly F1<F2<F3 in terms of the order of eating (with no difference in absolute choice in terms of numbers). From our first experiment we know that the dogs do not have a special preference for bread over chicken. So, the dogs are clearly following the gradient of chicken extract and ignoring the absolute protein content. Thus the dogs do seem to follow the “thumb rule” and must efficiently pick out food with higher concentrations of meat proteins going simply by the meat smell.

It, therefore, appears that these dogs have adapted to their scavenging habit without actually giving up the preference for meat. A possible mechanism might have been the development of better digestion of carbohydrates which has now been demonstrated to be one of the major genetic changes that the ancestors of dogs underwent during their transition from wolves (Axelsson et al. 2013). Given the carbohydrate rich diet of these dogs, this would be an advantage in terms of meeting their energy requirements, especially in areas like India where the human diet is chiefly comprised of carbohydrates. However, it seems that the dogs have behaviourally adapted to scavenging in human habitations by developing a thumb rule for foraging- “if it smells like meat, eat it”. This would enable them to always choose the food with a higher intensity of meat smell first, thus helping them sequester higher amounts of protein in their diet.

## Acknowledgements

This work was supported by grants from the Council for Scientific and Industrial Research, India and the Indian National Science Academy to AB, and by IISER-Kolkata. The experiments were designed by Anandarup Bhadra (ArB), and carried out by ArB, DB and MP. AB supervised the work and co-wrote the paper with ArB. Tithi Roy helped with the field work during these experiments. The authors would like to thank Dr. Ayan Banerjee, IISER-Kolkata for his comments and suggestions on the experiment and Prof. Raghavendra Gadagkar, Indian Institute of Science, Bangalore for his comments on the manuscript. We thank Dr. Tapas Kumar Sengupta, IISER-Kolkata and his students Paromita Banerjee and Brinta Chakraborty for their help in the biochemical analysis. Our experiments comply with the regulations for animal care in India.

**Legends to figures**

**Figure 1: Absolute choice data from experiment 1. B1 was chosen significantly more often than B2 or B3 (two tailed Fisher's Exact Test; p = 0.00), which were chosen equally as often (two tailed Fisher's Exact Test; p = 1).**

**Figure 2a: Known protein experiment with visual cues (KPEVC): Eating order is C>B>A (chi square = 45.37, df = 4, p = 0.00).**

**Figure 2b: Novel protein experiment with chicken smell (NPECS): eating oreder is T>I>S (chi square = 24.00, df = 4, p = 0.00).**

**Figure 2c: Known protein experiment with chicken smell (KPECS): Eating order is U>M>L (chi square = 11.14, df = 4, p = 0.02).**

**Figure 2d: Chicken and protein reverse gradient (CPRG): Eating order is F3>F2>F1 (chi square = 16.99, df = 4, p = 0.00)**

| Expt. No. | Expt. Name | Option 1 | # Times Selected | Option 2 | # Times Selected | Option 3 | # Times Selected |
|---|---|---|---|---|---|---|---|
| 1 | KPE | B1: Bread + Chicken extract | 28 | B2: Bread + Water | 7 | B3: Bread | 7 |
| 2 | KPEVC | A: Bread | 44 | B: Bread + curry (chicken) | 44 | C: Bread + Chicken in curry | 46 |
| 3A | NPECS | T: Pedigree soaked in 100% chicken extract | 27 | I: Pedigree soaked in 50% chicken extract | 25 | S: Pedigree soaked in water | 22 |
| 3B | KPECS | U: Bread + 100% Chicken extract | 26 | M: Bread + 50% Chicken extract | 26 | L: Bread + Water | 26 |
| 4 | CPRG | F1: 90 pellets of pedigree soaked in water | 14 | F2: 45 pellets of pedigree and half bread in 50% Chicken extract | 19 | F3: One and a half bread in 100% chicken extract | 21 |

**Table 1: Summary of experimental design and choices made in each experiment. KPE - Known Protein Experiment, KPEVC - Known Protein Experiment with Visual Cues (n = 46), NPECS - Novel Protein Experiment with Chicken Smell, KPECS - Known Protein Experiment with Chicken Smell, CPRG - Chicken and Protein Reverse Gradient (n = 30 for all other experiments).**

| Expt. No. | Chi-square value | P value for chi square | Log-likelihood value | P value for log-likelihood | Degrees of Freedom | Option chosen first (no. of times) | Option chosen second (no. of times) | Option chosen third (no. of times) |
|---|---|---|---|---|---|---|---|---|
| 2 | 45.371 | 0.000 | 42.875 | 0.000 | 4 | C(29) | B(26) | A(26) |
| 3a | 24.007 | 0.000 | 23.430 | 0.000 | 4 | T(18) | I(13) | S(13) |
| 3b | 11.139 | 0.025 | 11.570 | 0.021 | 4 | U(15) | M(10) | L(12) |
| 4 | 16.989 | 0.002 | 16.975 | 0.002 | 4 | F3(15) | F2(11) | F1(7) |

**Table 2: The results of the chi square tests performed to check for preference towards different food types provided in four experiments.**

**Figure 1**

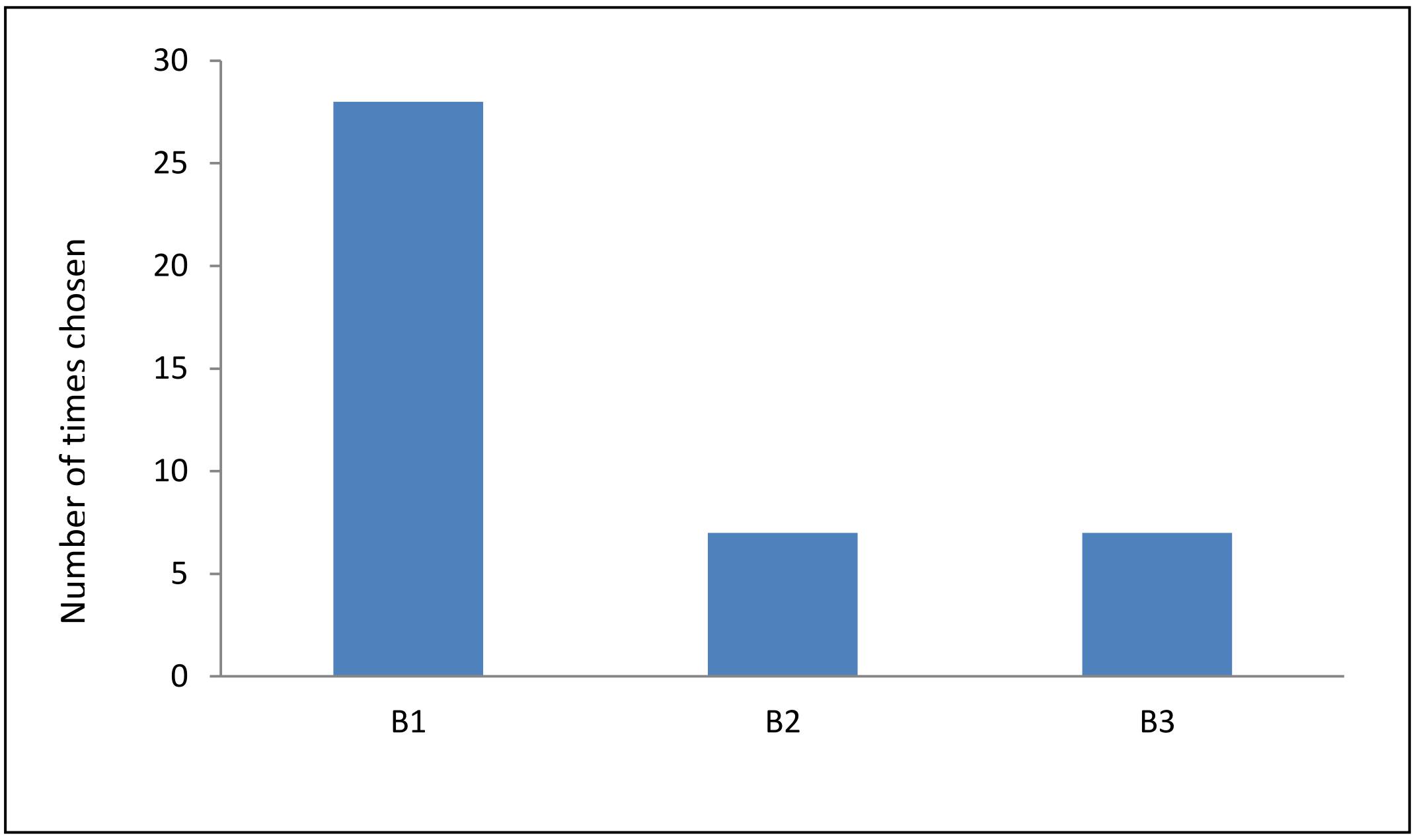

**Figure 2a**

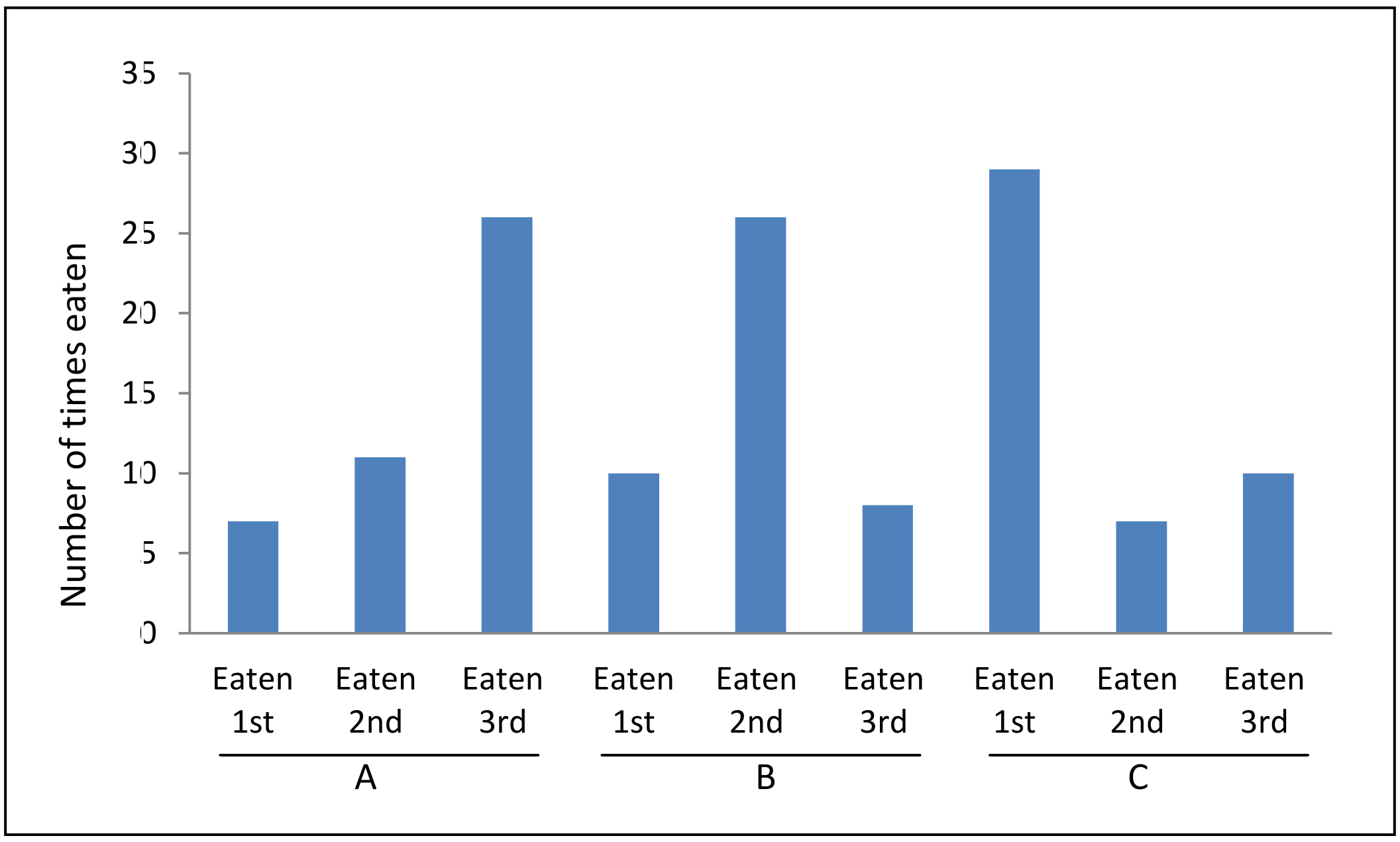

**Figure 2b**

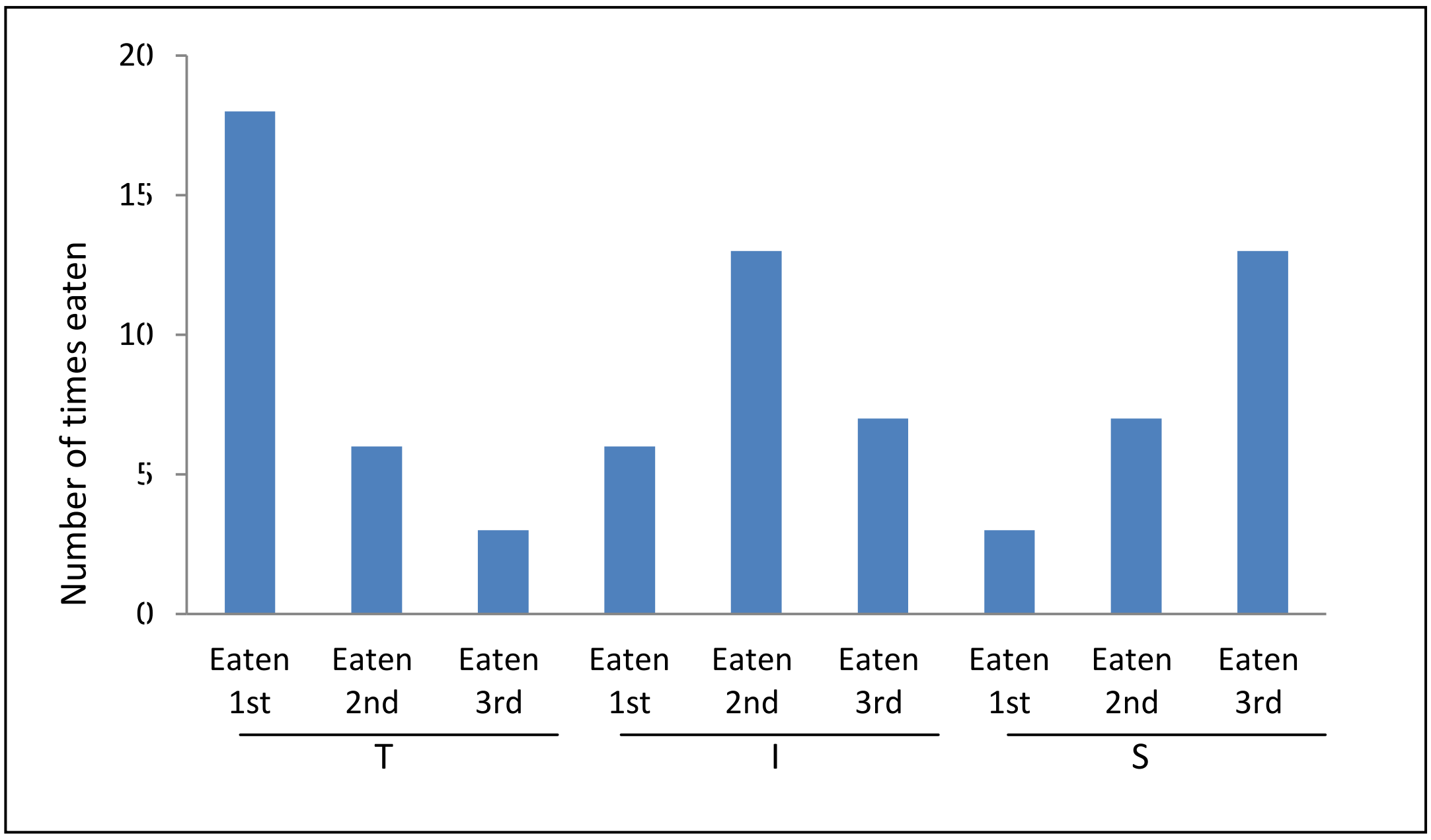

**Figure 2c**

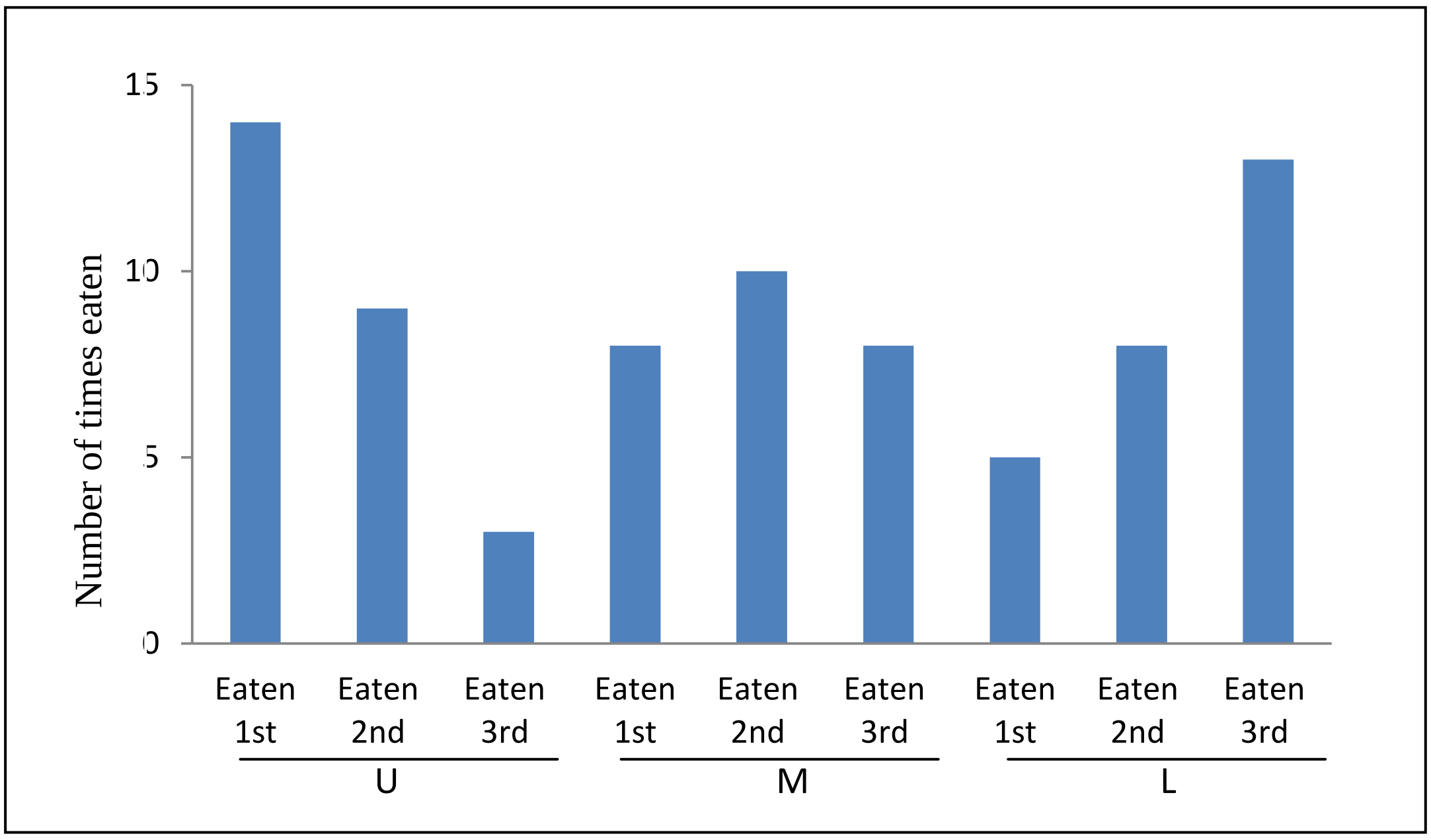

**Figure 2d**

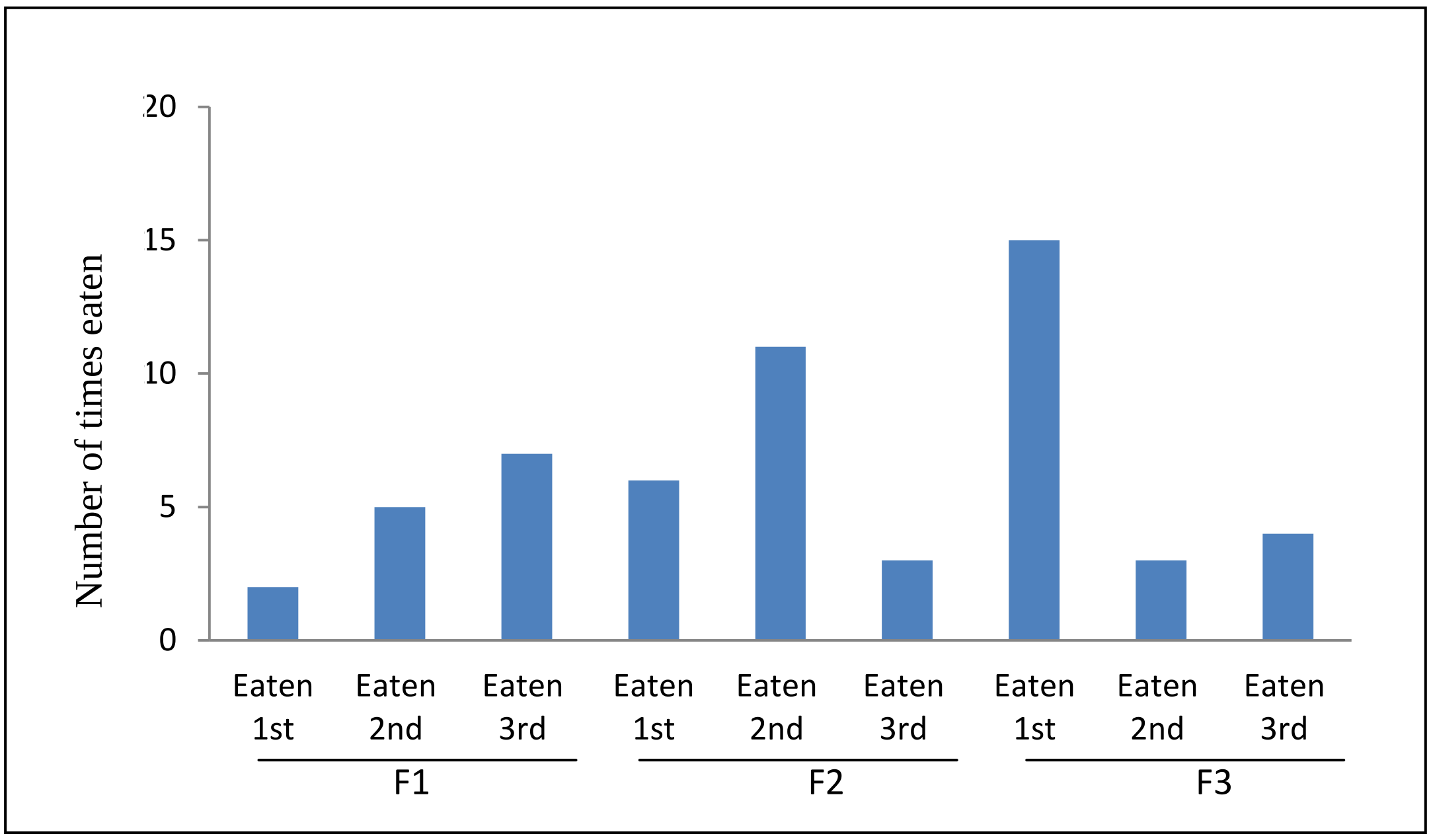

## Online Supplementary Material

### Preparation of chicken extract

- 50 gm of freshly cut chicken was put in 300 ml of commercially purified water and heated for 12 mins.
- The pieces of chicken and large particles were sieved off.
- The remaining liquid was allowed to cool. This final product was called chicken extract.

### Estimation of protein content in chicken extract

Protein content was estimated using Bradford method. Since particles that remain suspended in the extract may interfere with the Bradford method for protein estimation, we removed the small particles from the extract during the extraction and quantified the dissolved protein in the supernatant. Shimadzu UV-1800 was used for spectroscopy.

- A standard curve for light absorbance at λ595 was generated by taking O.D. readings for standard concentrations of BSA with commercially purified water as blank.
- 500 µl of freshly made chicken extract (was taken in three separate microcentrifuge tubes.
- Small particles were allowed to pellet down by spinning the tubes at 13.4 rpm for 2 mins.
- 2µl of the supernatant from each tube was added to 1ml Bradford reagent in a new microcentrifuge tube and mixed by turning over a few times.
- Samples were then incubated at 37°C for 15 mins.

- O.D. was checked at λ595 for each sample.
- Using the standard curve, the protein concentration was calculated for each sample.

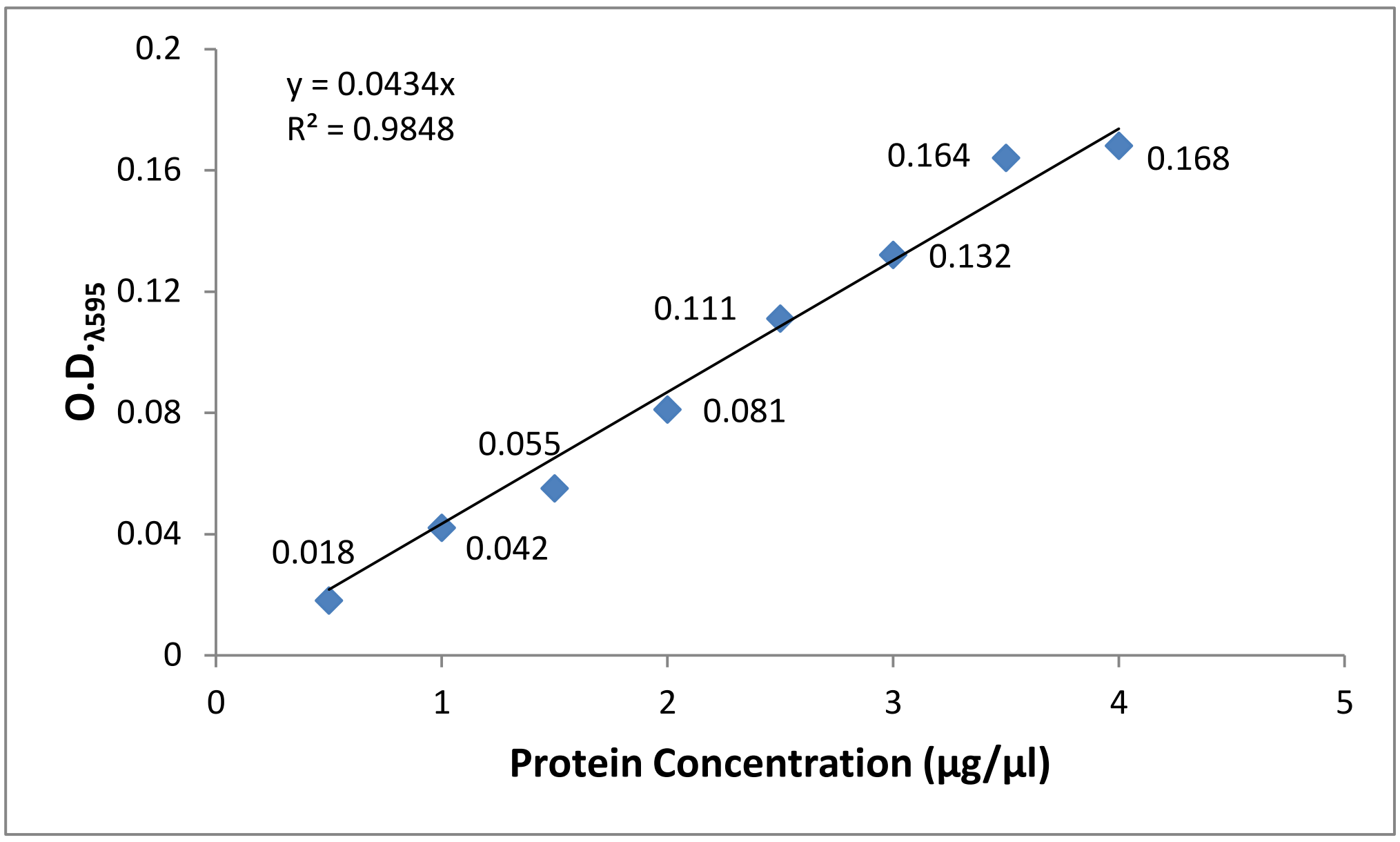

The three samples were found to have the following O.D.$_{595}$ readings: 0.058, 0.068 and 0.176.

So, protein concentrations are as follows: 1.35, 1.58 and 4.09 µg/2µl.

Therefore, 100 ml of the sample contains 0.067, 0.079 and 0.204 gm of protein.

Thus protein content in chicken extract is less than 0.25%(w/v)